\documentclass[11pt]{article}

\newcommand{\be}{\begin{equation}}
\newcommand{\ee}{\end{equation}}
\newcommand{\bea}{\begin{eqnarray}}
\newcommand{\eea}{\end{eqnarray}}
\newcommand{\ba}{\begin{array}}
\newcommand{\ea}{\end{array}}
\newcommand{\hone}{h_{1,1}}
\newcommand{\htwo}{h_{2,1}}
\newcommand{\M}{\mathcal{M}}
\newcommand{\N}{\mathcal{N}}

\newcommand{\R}{{\sf R\hspace*{-0.9ex}%
  \rule{0.15ex}{1.5ex}\hspace*{0.9ex}}}

\begin{document}

\thispagestyle{empty} \vspace{40pt} \hfill{hep-th/0610161}

\vspace{128pt}

\begin{center}
\textbf{\Large Wrapped M5-branes leading to \\ \vspace{10pt} five dimensional 2-branes }\\

\vspace{40pt}

Moataz H. Emam\footnote{Electronic address: memam@clarku.edu}

\vspace{12pt} \textit{Department of Physics}\\
\textit{Clark University}\\
\textit{Worcester, MA 01610}\\
\end{center}

\vspace{40pt}

\begin{abstract}

We study the dimensional reduction of M5-branes wrapping special
Lagrangian 3-cycles of a Calabi-Yau manifold and show explicitly
that they result in 2-branes coupled to the hypermultiplets of
ungauged $\mathcal{N}=2$ $D=5$ supergravity theory. In addition to
confirming previously known results, the calculation proves the
relationship between them and provides further insight on how the
topological properties of the compact space affect the lower
dimensional fields.

\end{abstract}

\newpage


\section{Introduction}

Black branes satisfying the Bogomol'nyi-Prasad-Sommerfield (BPS)
condition have been studied from a variety of perspectives many
times over the years, ever since it was discovered that such branes
preserve some degree of supersymmetry \cite{Gibbons:1982fy}. Of
particular interest are brane configurations wrapping manifolds with
special or restricted holonomy. Such manifolds have been catalogued
by Berger \cite{Berger} and shown to admit calibrated forms (an
excellent review is \cite{Joyce:2001nm}). This allows for the
construction of wrapped configurations simply by taking into account
calibrated forms on the compact space (see \cite{Emam:2004nc} or
\cite{Husain:2003ag} for detailed reviews and reference lists). Such
a program was started by the realization \cite{Cho:2000hg} that
certain constructions describing localized intersecting M5 branes
\cite{Fayyazuddin:1999zu} admit generalized K\"{a}hler calibrations,
which take into account the flux of the eleven dimensional 4-form
gauge field. This was further extended in the same reference
\cite{Cho:2000hg} and used to find more wrapped brane
configurations. Inevitably, branes wrapping other types of
calibrated cycles were sought after using a variety of techniques.
Of interest to us are branes wrapped over special
Lagrangian-calibrated (SLAG) submanifolds. The first such solution
was found in \cite{Emam:2004nc}, and later published in
\cite{Emam:2005bh}. A more general SLAG calibrated construction was
announced in \cite{Martelli:2003ki}. In the same reference, branes
wrapping submanifolds with $G_2$ holonomy were also studied. Later,
Fayyazuddin and Husain \cite{Fayyauddin:2005as} took a second look
at SLAG wrapped branes and proceeded to explore other instances of
SLAG wrappings \cite{Fayyazuddin:2005pm, Fayyazuddin:2005ds}.

On the other hand, it is also possible to argue that M-branes
wrapping calibrated cycles of a Calabi-Yau manifold dimensionally
reduce to interesting BPS configurations in lower dimensions. This
has always been an assumption based on geometric and physical
arguments. For example, the K\"{a}hler-calibrated branes of
\cite{Cho:2000hg} dimensionally reduce to black holes and strings
coupled to the vector multiplets of $\mathcal{N}=2$ $D=5$
supergravity as demonstrated in \cite{Kastor:2003jy}. In addition to
being formal proofs of hypotheses that have always been simply
assumed, such calculations provide deeper insights into how lower
dimensional fields arise as consequences of the topology of the
compact subspace.

The other fields sector of ungauged $D=5$ supergravity theory is the
hypermultiplets sector. A particular special case of that is the
so-called universal hypermultiplet. This is better understood from a
higher dimensional viewpoint as the dimensional reduction over SLAG
cycles of a Calabi-Yau submanifold with constant complex structure
moduli (a lightning review is presented in the next section). In
\cite{Emam:2005bh}, we found explicit SLAG-wrapped solutions in
$D=11$ and showed how such wrappings dimensionally reduce to
2-branes in $D=5$ coupled only to the universal hypermultiplet
fields. In the same paper we analyzed the conditions on a more
general $D=5$ 2-brane that couples to the full set of
hypermultiplets. From the higher dimensional perspective, this is
assumed to be a M5-brane wrapping SLAG cycles of a Calabi-Yau 3-fold
with non-trivial complex structure moduli; the same calibrated brane
configuration studied by \cite{Martelli:2003ki} and
\cite{Fayyauddin:2005as}. This correspondence has been assumed
without proof in most of these references. In this paper, we provide
that proof and show that at a scale much larger than the size of the
compact space, one does indeed retrieve the results of
\cite{Emam:2005bh}.

The paper is structured as follows: Section \ref{compactification}
reviews how $D=11$ supergravity reduces to five dimensions,
producing the ungauged $\mathcal{N}=2$ theory, and sets the
conventions and notation. We are only interested in the
hypermultiplets, so the vector multiplets sector will be ignored.
Section \ref{thesolutions} presents the eleven dimensional
SLAG-wrapped M5-brane in the form of \cite{Fayyauddin:2005as} as
well as the 2-brane of \cite{Emam:2005bh}. At the risk of confusing
the reader, we had to change the notations used in the two
references slightly in order to avoid using the same symbols to
describe different things. Section \ref{thecalculation} details the
dimensional reduction of the eleven dimensional brane to produce the
five dimensional result exactly and analyzes the geometric meaning
of the $D=11$ SUSY conditions in the far field limit. Finally we
conclude and propose further study.

\section{Dimensional reduction of $D=11$ supergravity}\label{compactification}

Dimensionally reducing $D=11$ supergravity on a Calabi-Yau $3$-fold
$\M$ yields ungauged $D=5$ $\mathcal{N}=2$ supergravity coupled to
$(\hone-1)$ vector multiplets and $(\htwo+1)$ hypermultiplets
\cite{Cadavid:1995bk}; the $h$'s being the Hodge numbers of $\M$.
M-branes wrapping K\"{a}hler calibrated cycles of $\M$ deform the
K\"{a}hler structure of $\M$ and reduce to configurations in which
the vector multiplets are excited \cite{Kastor:2003jy}. SLAG wrapped
M-branes, our focus in this paper, deform the complex structure of
$\M$ and reduce to configurations carrying charge under the
hypermultiplet scalars. The two sectors of the theory decouple and
we only keep the hypermultiplets in our presentation.

The $D=5$ $\mathcal{N}=2$ supergravity Langrangian including the
full set of $(\htwo+1)$ hypermultiplets can be written in terms of
geometric quantities on the moduli space of the complex structures
of the Calabi-Yau manifold $\M$.  These structures are discussed
in detail in \cite{Candelas:1990pi} and we will give a brief
review here. Start by taking a basis of the homology $3$-cycles
$(A^I,B_J)$ with $I,J=0,1,\dots,\htwo$ and a dual cohomology basis
of $3$-forms $(\alpha_I,\beta^J)$ such that
\begin{equation}
\int_{A^J}\alpha_I=\int_{\M}\alpha_I\wedge\beta^J=\delta_I^J,\qquad
\int_{B_I}\beta^J=\int_{\M}\beta^J\wedge\alpha_I=-\delta_I^J.
\end{equation}
Define the periods of the holomorphic $3$-form $\Omega$ on $\M$ by
\begin{equation}\label{Omega}
Z^I=\int_{A^I}\Omega,\qquad   F_I=\int_{B_I}\Omega.
\end{equation}

The periods $Z^I$ can be regarded as coordinates on the complex
structure moduli space.  Since $\Omega$ can be multiplied by an
arbitrary complex number without changing the complex structure,
the $Z^I$ are projective coordinates.  The remaining periods $F_I$
can then be regarded as functions $F_I(Z)$. One can further show
that $F_I$ is the gradient of a function $F(Z)$, known as the
prepotential, that is homogeneous of degree two in the
coordinates, {\it i.e.} $F_I=\partial_I F(Z)$ with $F(\lambda
Z)=\lambda^2 F(Z)$. The quantity $F_{IJ}(Z)=\partial_I\partial_J
F(Z)$ will also play an important role. Non-projective coordinates
can then be given by taking {\it e.g.} $z^i=Z^i/Z^0$ with
$i=1,\dots,\htwo$. The K\"{a}hler potential of the complex
structure moduli space is $\mathcal{K}=-\ln
(i\int_\M\Omega\wedge\bar\Omega)$. Given the expansion of $\Omega$
in terms of the periods
\begin{equation}
\Omega=Z^I\alpha_I-F_I\beta^I, \label{OmegaDef}
\end{equation}
the K\"{a}hler potential is determined in terms of the
prepotential $F(Z)$ according to
\begin{equation}
\mathcal{K} = -\ln\left[ i ( Z^I \bar F_I - \bar Z^I F_I
)\right].\label{KahlerPotential}
\end{equation}

The so-called period matrix $\N_{IJ}$ is defined by
\begin{equation}
\N_{IJ} = \bar F_{IJ} -2 i {N_{IK}Z^KN_{JL}Z^L\over Z^P N_{PQ}
Z^Q}    = \theta_{IJ}-i \gamma_{IJ}\label{period2}
\end{equation}
where $N_{IJ}=Im(F_{IJ})$, $\gamma^{IJ}\gamma_{JK}=\delta^I_K$,
and $\left( \theta, \gamma \right )$ are real matrices.

The derivation of the Lagrangian for the bosonic fields of the
$D=5$ theory is sketched in \cite{Gutperle:2000ve} and detailed in
\cite{Emam:2004nc}. The bosonic part of the $D=11$ action is the
familiar:
\begin{equation}\label{eleven}
    S_{11}  = \frac{1}{{2\kappa _{11}^2 }}\int {d^{11} x\sqrt { - G}
    \left( {R - \frac{1}{{48}}F^2 } \right)}  - \frac{1}{12\kappa _{11}^2 } \int {A \wedge F \wedge
    F},
\end{equation}
where $F$ is given by $F=dA$; $A$ being the usual eleven dimensional
3-form gauge field. The dimensional reduction of (\ref{eleven}) is
done over the metric:
\begin{eqnarray}
    ds^2  &=&   e^{2\sigma/3}g_{\mu\nu}dx^\mu
    dx^\nu+e^{-\sigma/3}k_{\tilde I \tilde J} dx^{\tilde I} dx^{\tilde J} ,\nonumber\\\mu,\nu&=&0,\ldots,4\quad\quad \tilde I, \tilde J=1,\ldots ,6  \label{wrappingmetric}
\end{eqnarray}
where $k_{\tilde I \tilde J}$ is a fixed Ricci flat metric on the
Calabi-Yau space $\M$. The $D=11$ 3-form is expanded in terms of
the cohomology basis as follows
\begin{equation}
    A = {1\over 3!}A_{\mu\nu\rho} dx^\mu\wedge dx^\nu\wedge dx^\rho + \sqrt 2 \left( {\zeta
^I \alpha _I  + \tilde \zeta _I \beta ^I }
    \right),
\end{equation}
and
\begin{eqnarray}
 F = dA &=& \frac{1}{{4!}}F_{\mu \nu \rho \sigma } dx^\mu   \wedge dx^\nu   \wedge dx^\rho   \wedge dx^\sigma   \nonumber\\
  &+& \sqrt 2 \left[ {\left( {\partial _\mu \zeta ^I } \right)\alpha _I  + \left( {\partial _\mu \tilde \zeta _I } \right)\beta ^I } \right] \wedge
  dx^\mu. \label{FExpanded}
\end{eqnarray}

The resulting $D=5$ bosonic action is
\begin{eqnarray}
    S_5  &=& \frac{1}{{2\kappa _{5}^2 }}\int {d^5 x\sqrt { - g} \left[ { R - \frac{1}{2}\left( {\partial
    _\mu \sigma     } \right)\left( {\partial ^\mu  \sigma } \right) - G_{i\bar j} (
    {\partial _\mu  z^i } )( {\partial ^\mu  z^{\bar j} } )}
    \right.}  \nonumber \\
    &-& \left.{ \frac{1}{{48}}e^{ - 2\sigma } F_{\mu \nu \rho \sigma } F^{\mu \nu \rho \sigma }-\frac{1}{{24}}\varepsilon _{\mu \nu \rho \sigma \alpha } F^{\mu \nu
    \rho         \sigma } K^\alpha  \left( {\zeta ,\tilde \zeta } \right) + e^\sigma
    L_\mu     ^\mu  \left( {\zeta ,\tilde \zeta } \right)} \right], \label{d5theory}
\end{eqnarray}
where we have defined:
\begin{eqnarray}
    K_\alpha  ( {\zeta ,\tilde \zeta } ) &=& \left[ {\zeta ^I (
    {\partial_\alpha  \tilde \zeta _I } ) - \tilde \zeta _I ( {\partial_\alpha  \zeta
    ^I     } )} \right] \nonumber \\
    L_{\mu \nu}  ( {\zeta ,\tilde \zeta } ) &=&  - \left( {{\gamma _{IJ}  + \gamma ^{KL} \theta _{IK} \theta _{JL} } } \right)\left(
    {\partial _\mu  \zeta^I } \right)\left( {\partial _\nu  \zeta^J } \right)
    - \gamma ^{IJ} ( {\partial _\mu  \tilde \zeta_I } )( {\partial _\nu
    \tilde \zeta_J     } ) \nonumber \\
    & & -2\gamma ^{IK} \theta _{KJ}  \left( {\partial _\mu  \zeta^J } \right)(
    {\partial _\nu  \tilde \zeta_I } ). \label{KandL}
\end{eqnarray}

The scalar fields $z^i$, $z^ {\bar i}$ with $i=1,\dots,\htwo$ are
complex coordinates on the complex structure moduli space with metric
$G_{i\bar j} = \partial _i \partial _{\bar j} \mathcal{K}$. The
pseudo-scalar axions $\left(\zeta^I,\tilde\zeta_I\right)$ arise from
the dimensional reduction of the $D=11$ $3$-form gauge potential. The
scalar field $\sigma$ is the overall volume scalar of $\M$ and
$F_{\mu \nu \rho \sigma }$ is the $D=5$ 4-form field strength. Each
hypermultiplet has 4 scalar fields. The scalars $(z^i,z^ {\bar
i},\zeta^i,\tilde \zeta_i)$ make up $\htwo$ of the hypermultiplets,
while the additional universal hypermultiplet is comprised of the
fields $(a,\sigma,\zeta^0,\tilde \zeta_0)$, where the so-called
universal axion $a$ is the scalar dual to the 3-form gauge potential
$A_{\mu\nu\rho}$.

Further study of the structure of the theory (see
\cite{Gutperle:2000ve} and the references within) reveals that the
hypermultiplets define a $(\htwo+1)$-dimensional quaternionic
space. This structure, in five dimensions, is dual to the special
K\"{a}hler geometry of the $D=4$ vector multiplets sector via the
so called c-map ({\it e.g.} \cite{DeJaegher:1997ka}). This duality
justifies the use of special K\"{a}hler geometry techniques as
opposed to the explicit quaternionic form.

Furthermore, one finds that the theory is invariant under the
symplectic group $Sp(2\htwo,\R)$, {\it i.e.} (\ref{d5theory})
actually defines a family of Lagrangians that differ from each
other only by a rotation in symplectic space that has no effect on
the physics. In fact, if we define
\begin{equation}\label{symvec}
    V = \left( {\begin{array}{*{20}c}
       {L^I }  \\
       {M_J }  \\
    \end{array}} \right) \equiv e^{{\mathcal{K} \mathord{\left/
     {\vphantom {\mathcal{K} 2}} \right.
     \kern-\nulldelimiterspace} 2}} \left( {\begin{array}{*{20}c}
       {Z^I }  \\
       {F_J }  \\
    \end{array}} \right)
\end{equation}
satisfying
\begin{equation}\label{covderiv}
    \nabla _{\bar i} V = \left[ {\partial _{\bar i}  -
    \frac{1}{2}\left( {\partial _{\bar i} \mathcal{K}} \right)} \right]V =
    0,
\end{equation}
then $V$ is a basis vector in symplectic space that satisfies the
inner product
\begin{equation}\label{innerprod}
    i\left\langle V \right|\left. {\bar V} \right\rangle  = i\left(
    {\bar L^I M_I  - L^I \bar M_I } \right) = 1.
\end{equation}
An orthogonal vector may be defined by
\begin{equation}\label{covderiv1}
    U_i  \equiv \left(
    {\begin{array}{*{20}c}
       {f_i^I }  \\
       {h_{J|i} }  \\
    \end{array}} \right)=\nabla _i V,
\end{equation}
such that
\begin{equation}
    \left\langle V \right|\left. {U_i } \right\rangle  = \left\langle V \right|\left. {U_{\bar i} } \right\rangle  = 0.
\end{equation}

Based on this, the following identities may be derived:
\begin{eqnarray}
    \mathcal{N}_{IJ} L^J  &=& M_I \quad ,\quad \mathcal{N}_{IJ} f_i^J  =
    h_{I|i}\label{comp1}\\
    \left( {\nabla _{\bar j} f_i^I } \right) &=& G_{i\bar j} L^I ,\quad \left( {
\nabla _{\bar j} h_{iI} } \right) = G_{i\bar j} M_I
    \label{comp2} \\
    \gamma _{IJ} L^I \bar L^J  &=&   \frac{1}{2}, \quad\quad G_{i\bar j}  =    2f_i^I \gamma _{IJ} f_{\bar
    j}^J,\label{comp3}
\end{eqnarray}
as well as the very useful:
\begin{eqnarray}
 \gamma ^{IJ}  &=& 2\left( {G^{i\bar j} f_i^I f_{\bar j}^J  + L^I \bar L^J } \right) \label{comp4}\\
 \left( {\gamma _{IJ}  + \gamma ^{KL} \theta _{IK} \theta _{JL} } \right) &=& 2\left( {G^{i\bar j} h_{iI} h_{\bar jJ}  + M_I \bar M_J } \right) \label{comp5}\\
 \gamma ^{IK} \theta _{KJ}  &=& 2\left( {L^I \bar M_J  + \bar L^I M_J }
 \right).\label{comp6}
\end{eqnarray}

Such detail follows directly from the topology of the underlying
compact manifold, and it is indeed a wonder that we can understand
so much about it with little need for the explicit form of a
metric on $\M$.

\section{Wrapped M5-branes and $D=5$ 2-branes with
hypermultiplets}\label{thesolutions}

The proposition we are attempting to analyze in this paper is that
M5-branes wrapping SLAG cycles of a CY 3-fold dimensionally reduce
to 2-branes coupled to the hypermultiplets of $\mathcal{N}=2$ $D=5$
ungauged SUGRA. In this section we summarize both constructions as
they were presented in references \cite{Fayyauddin:2005as} and
\cite{Emam:2005bh}.

\subsection{The $D=5$ 2-brane}\label{2brane}

Based on the notation established in \S\ref{compactification}, the
$D=5$ 2-brane spacetime metric coupled to the hypermultiplets may be
written as follows \cite{Emam:2005bh}:
\begin{equation}\label{D5metric}
    ds^2  = (-dt^2+dx_1^2+dx_2^2) + e^{-2\sigma } \delta _{ab }
    dx^a   dx^b,\quad \quad a,b=3,4. \\
\end{equation}

We define a number $\left ( h_{2,1}+1 \right )$ of harmonic
functions
\begin{equation}
    H_I  = h_I  + q_I \ln r\quad ,\quad \tilde H^I  = \tilde h^I  + \tilde q^I \ln
    r \quad ,\quad I = 0, \ldots ,h_{2,1},
\end{equation}
where $h$ and $\tilde h$ are constants, $r$ is the radial coordinate
in the two dimensional space transverse to the brane and $(q,\tilde
q)$ are electric and magnetic charges. The SUSY equations yield the
following constraints on the scalar fields:
\begin{eqnarray}
    \left( {\partial _a  \sigma } \right) &=&  - 2e^{ \frac{\sigma}{2} } \left
[
    {L^I     \left( {\partial _a  H_I } \right) - M_I \left( {\partial _a  \tilde H^I
    }     \right)} \right] \nonumber \\
&=&  - 2e^{ \frac{\sigma}{2} } \left [
    {\bar L^I     \left( {\partial _a  H_I } \right) - \bar M_I \left( {\partial _a  \tilde H^I
    }     \right)} \right] \label{sigma}\\
    \left( {\partial _a  z^i } \right) &=& - e^{ \frac{\sigma}{2} } G^{i\bar j
}
    \left[     {f_{\bar j}^I \left( {\partial _a  H_I } \right) - h_{\bar jI}
\left(
    {\partial _a  \tilde H^I } \right)} \right] \nonumber \\
    \left( {\partial _a  z^{\bar i} } \right) &=& -e^{ \frac{\sigma}{2} } G^{\bar ij}    \left[ {f_j^I \left( {\partial _a  H_I } \right) - h_{jI} \left( {\partial
    _a    \tilde H^I } \right)} \right] \label{zzzs}\\
    \left( {\partial _a  \zeta ^I } \right) &=& \pm
    {\varepsilon_a}^{\;c}   \left( {\partial
    _c      \tilde H^I } \right)\nonumber \\
    \left( {\partial _a  \tilde \zeta _I } \right) &=& \pm
    {\varepsilon_a}^{\;c}  \left( {\partial _c  H_I } \right), \label{zetaeom}
\end{eqnarray}
and $F_{\mu \nu \rho \sigma}=0$. Using a well-known relationship
between the charges $q$ and $\tilde q$ and the central charge $Z$
of the theory as follows \cite{Witten:mh}:
\begin{eqnarray}
    Z &=& \left( {L^I q_I  - M_I \tilde q^I } \right) \nonumber \\
    \bar Z &=& \left( {\bar L^I q_I  - \bar M_I \tilde q^I }
    \right),
\end{eqnarray}
equations (\ref{sigma},\ref{zzzs}) may be simplified to:
\begin{eqnarray}
    \frac{{d\sigma }}{{dr}} &=&  - 2e^{  \frac{\sigma}{2} } \frac{Z}{r}\nonumber \\
    \frac{{dz^i }}{{dr}} &=&  - e^{ \frac{\sigma}{2} } \frac{{\nabla ^i \bar Z}}
{r}
    \nonumber \\
    \frac{{dz^{\bar i} }}{{dr}} &=&  - e^{ \frac{\sigma}{2} } \frac{{\nabla^{\bar i}
    Z}}{r},     \label{zzeom}
\end{eqnarray}
which may further be shown to satisfy \cite{Sabra:1997dh}:
\begin{equation}
    H_I  = i\left( {F_I  - \bar F_I } \right)\quad \quad \tilde H^I =
    i( {Z^I  - \bar Z^I } ).
\end{equation}

\subsection{The wrapped M5-brane}\label{M5brane}

The spacetime metric derived by Fayyazuddin and Husain in reference
\cite{Fayyauddin:2005as} may be written as follows:

\begin{eqnarray}
    ds^2  &=& H^{ - {1 \mathord{\left/
 {\vphantom {1 3}} \right.
 \kern-\nulldelimiterspace} 3}} \left( { - dt^2  + dx_1^2  + dx_2^2 } \right) + g_{\tilde I\tilde J} dx^{\tilde I} dx^{\tilde J}  + H^{{2 \mathord{\left/
 {\vphantom {2 3}} \right.
 \kern-\nulldelimiterspace} 3}} \delta _{ab} dx^a dx^b \nonumber\\ \tilde I,\tilde J &=& 1, \ldots ,6\,\,\,\,\,\,\,\,a,b =
 3,4, \label{11Dmetric}
\end{eqnarray}
representing a M5-brane wrapping a Calabi-Yau 3-fold with metric
$g_{\tilde I\tilde J}$. The brane's tension distorts the compact
space such that it is no longer strictly Calabi-Yau. The scale
factor $H$ is a function in the two dimensional transverse space.

The brane is naturally coupled to the eleven dimensional 7-form
field strength which was constructed in the same reference. For
our purposes, its dual 4-form field strength (which they also
gave) may be more useful. It is:
\begin{eqnarray}
    F_{(4)}  &=& \frac{1}{4}H^{{1 \mathord{\left/
 {\vphantom {1 6}} \right.
 \kern-\nulldelimiterspace} 6}} \varepsilon _{ab}  \star _6 d_6 \left[ {H^{{1 \mathord{\left/
 {\vphantom {1 2}} \right.
 \kern-\nulldelimiterspace} 2}} \left( {{\mathop{\rm Re}\nolimits} \Omega  } \right)} \right]  \wedge dx^a  \wedge dx^b\nonumber\\
   &-& \frac{i}{2}H^{ - {1 \mathord{\left/
 {\vphantom {1 2}} \right.
 \kern-\nulldelimiterspace} 2}}  \varepsilon _a ^{\,\,\,\,\, b} \partial _b \left[ {H^{{1 \mathord{\left/
 {\vphantom {1 2}} \right.
 \kern-\nulldelimiterspace} 2}} \left( {{\mathop{\rm Im}\nolimits} \Omega  } \right)} \right] \wedge
 dx^a, \label{Field4}
\end{eqnarray}
where $\Omega$ is a globally defined holomorphic form which turns
out to be the usual Calabi-Yau 3-form and $\star_6$ is the Hodge
dual operator on $\M$.

Dictated by SUSY preservation, certain constraints on the compact
manifold were also found:
\begin{eqnarray}
\bar \Omega   \wedge  \star _6 d_6 \Omega   &=& 0 \label{const3}\\
 d_6 \left( {\Omega   - \bar \Omega  } \right) &=&
 0\label{const4}\\
 \det \left( {g_{\tilde I\tilde J} } \right) \equiv g &=& H \label{const1}\\
 \Omega ^{\tilde I\tilde J\tilde K} \left ( \partial _a \bar \Omega  _{\tilde I\tilde J\tilde K}\right )  &=& 12\partial _a \ln g. \label{const2}
\end{eqnarray}

\section{Dimensional reduction and analysis}\label{thecalculation}

We now show that the $D=5$ 2-brane of \S\ref{2brane} is the
dimensional reduction of the SLAG-wrapped M5-brane of
\S\ref{M5brane}. Since we already have a dimensional reduction
scheme, as given in \S \ref{compactification}, one can simply
merge the $D=11$ equations into that scheme and see if what we
retrieve is consistent with the $D=5$ results.

We begin by considering the metric (\ref{11Dmetric}). Rearranging:
\begin{equation}
ds^2  = H^{ - {1 \mathord{\left/
 {\vphantom {1 3}} \right.
 \kern-\nulldelimiterspace} 3}} \left( { - dt^2  + dx_1^2  + dx_2^2  + H\delta _{ab} dx^a dx^b }
 \right)+ g_{\tilde I\tilde J} dx^{\tilde I} dx^{\tilde J},
\end{equation}
and comparing with the form of the metric (\ref{wrappingmetric})
used for the dimensional reduction of the theory, one is forced to
conclude that:
\begin{equation}\label{determinant}
H = e^{ - 2\sigma } ,\,\,\,\,\,\,\,\,\,\,\,\,\,\,g_{\tilde I\tilde
J}  = H^{{1 \mathord{\left/
 {\vphantom {1 6}} \right.
 \kern-\nulldelimiterspace} 6}} k_{\tilde I\tilde J}  = e^{ - {\sigma  \mathord{\left/
 {\vphantom {\sigma  3}} \right.
 \kern-\nulldelimiterspace} 3}} k_{\tilde I\tilde J}
\end{equation}

Based on this, one immediately sees that the five dimensional
metric is:
\begin{eqnarray}
 g_{\mu \nu } dx^\mu  dx^\nu   &=&  \left ( - dt^2  + dx_1^2  + dx_2^2\right )  + H\delta _{ab} dx^a dx^b  \nonumber\\
  &=&  \left ( - dt^2  + dx_1^2  + dx_2^2\right )  + e^{ - 2\sigma } \delta _{ab} dx^a
  dx^b,
\end{eqnarray}
which is exactly the $D=5$ result (\ref{D5metric}).

Next, we turn to the field strength. To begin with, we argue that as
the compact subspace is shrunk to a point, variations on $\M$ vanish
and expressions such as $d_6\left [ H^{{1 \mathord{\left/ {\vphantom
{1 2}} \right.\kern-\nulldelimiterspace} 2}} \left ( {\mathop{\rm
Re}\nolimits} \Omega \right ) \right ]$ can be neglected, so we set
the first term of (\ref{Field4}) to zero. We also identify $\Omega$
as the holomorphic Calabi-Yau 3-form. To facilitate the calculation,
we make the assumption that $\sigma  = \mathcal{K}$, where the
K\"{a}hler potential $\mathcal{K}$ is defined by
(\ref{KahlerPotential})\footnote{This is essentially the assumption
made by \cite{Sabra:1997dh} in the context of solving the attractor
equations of four dimensional black holes coupled to the vector
multiplets.}, and make use of (\ref{OmegaDef}) as follows:
\begin{eqnarray}
 \Omega  &=& \left( {Z^I \alpha _I  - F_I \beta ^I } \right) = e^{ - {\sigma  \mathord{\left/
 {\vphantom {\sigma  2}} \right.
 \kern-\nulldelimiterspace} 2}} \left( {L^I \alpha _I  - M_I \beta ^I } \right)\,\,{\rm and}\,\,c.c. \nonumber\\
 \left ( \partial _a \Omega\right )  &=& e^{ - {\sigma  \mathord{\left/
 {\vphantom {\sigma  2}} \right.
 \kern-\nulldelimiterspace} 2}} \left[ {\left( {\partial _a L^I } \right)\alpha _I  - \left( {\partial _a M_I } \right)\beta ^I } \right] - \frac{1}{2}e^{ - {\sigma  \mathord{\left/
 {\vphantom {\sigma  2}} \right.
 \kern-\nulldelimiterspace} 2}} \left( {\partial _a \sigma } \right)\left( {L^I \alpha _I  - M_I \beta ^I } \right)\,\,{\rm
 and}\,\,c.c.,
\end{eqnarray}
where $c.c.$ means the complex conjugate of each of these equations.
We find:
\begin{eqnarray}
 F &=&  - \frac{i}{4} \varepsilon _a^{\,\,\,\,\,b} \left[ {\left( {\partial _b \Omega } \right) - \left( {\partial _b \bar \Omega } \right)} \right]\wedge dx^a + \frac{i}{4}\varepsilon _a^{\,\,\,\,\,b} \left( {\partial _b \sigma } \right)\left[ {\Omega  - \bar \Omega } \right] \wedge dx^a\\
  &=& \frac{i}{4} \varepsilon _a^{\,\,\,\,\,b} e^{ - {\sigma  \mathord{\left/
 {\vphantom {\sigma  2}} \right.
 \kern-\nulldelimiterspace} 2}} \left[ { - \left( {\partial _b L^I } \right) + \left( {\partial _b \bar L^I } \right) - \frac{1}{2}\left( {\partial _b \sigma } \right)L^I  + \frac{1}{2}\left( {\partial _b \sigma } \right)\bar L^I } \right]\alpha _I  \wedge dx^a  \nonumber\\
  &+& \frac{i}{4} \varepsilon _a^{\,\,\,\,\,b} e^{ - {\sigma  \mathord{\left/
 {\vphantom {\sigma  2}} \right.
 \kern-\nulldelimiterspace} 2}} \left[ {\left( {\partial _b M_I } \right) - \left( {\partial _b \bar M_I } \right) + \frac{1}{2}\left( {\partial _b \sigma } \right)M_I  - \frac{1}{2}\left( {\partial _b \sigma } \right)\bar M_I } \right]\beta ^I  \wedge
 dx^a. \label{MidCalc1}
\end{eqnarray}

It is straightforward to show that \cite{Emam:2004nc}:
\begin{equation}
    \left( {\partial _b L^I } \right) = f_i^I \left( {\partial _b z^i } \right),\,\,\,\,\left( {\partial _b M_I } \right) = h_{iI} \left( {\partial _b z^i } \right)\,\,{\rm
    and}\,\,c.c.
\end{equation}

For vanishing universal axion, equation (\ref{FExpanded}) becomes
\begin{equation}
 F = \sqrt 2 \left[ {\left( {\partial _a \zeta ^I } \right)\alpha _I  + \left( {\partial _a \tilde \zeta _I } \right)\beta ^I } \right] \wedge
  dx^a,
\end{equation}
which we compare with (\ref{MidCalc1}) to conclude:
\begin{eqnarray}
 \left( {\partial _a \zeta ^I } \right) &=& \frac{i}{{4\sqrt 2 }} \varepsilon _a^{\,\,\,\,\,b} e^{ - {\sigma  \mathord{\left/
 {\vphantom {\sigma  2}} \right.
 \kern-\nulldelimiterspace} 2}} \left[ { - f_i^I \left( {\partial _b z^i } \right) + f_{\bar i}^I \left( {\partial _b z^{\bar i} } \right) - \frac{1}{2}\left( {\partial _b \sigma } \right)L^I  + \frac{1}{2}\left( {\partial _b \sigma } \right)\bar L^I } \right] \label{MidCalc3}\\
 \left( {\partial _a \tilde \zeta _I } \right) &=& \frac{i}{{4\sqrt 2 }} \varepsilon _a^{\,\,\,\,\,b} e^{ - {\sigma  \mathord{\left/
 {\vphantom {\sigma  2}} \right.
 \kern-\nulldelimiterspace} 2}} \left[ {h_{iI} \left( {\partial _b z^i } \right) - h_{\bar iI} \left( {\partial _b z^{\bar i} } \right) + \frac{1}{2}\left( {\partial _b \sigma } \right)M_I  - \frac{1}{2}\left( {\partial _b \sigma } \right)\bar M_I }
 \right].\label{MidCalc2}
\end{eqnarray}

The crucial step is to insert the constraints (\ref{sigma}) and
(\ref{zzzs}) in their respective slots in
(\ref{MidCalc3},\ref{MidCalc2}) and see if we can retrieve
(\ref{zetaeom}). Doing so for (\ref{MidCalc3}) and rearranging:
\begin{eqnarray}
 \left( {\partial _a \zeta ^I } \right) &=& \frac{i}{{4\sqrt 2 }} \varepsilon _a^{\,\,\,\,\,b} \left[ {\left( {f_i^I f_{\bar j}^J G^{i\bar j}  + L^I \bar L^J } \right) - \left( {f_j^J f_{\bar i}^I G^{j\bar i}  + L^J \bar L^I } \right)} \right]\left( {\partial _b H_J } \right) \nonumber\\
  &+& \frac{i}{{4\sqrt 2 }} \varepsilon _a^{\,\,\,\,\,b} \left[ { - f_i^I h_{\bar jJ} G^{i\bar j}  - L^I \bar M_J  + f_{\bar i}^I h_{jJ} G^{\bar ij}  + \bar L^I M_J } \right]\left( {\partial _b \tilde H^J }
  \right).
\end{eqnarray}

The first term cancels out, while for the second we can use the
identities (\ref{comp1}) as well as the definition (\ref{period2})
to get:
\begin{eqnarray}
 \left( {\partial _a \zeta ^I } \right) &=& \frac{i}{{4\sqrt 2 }} \varepsilon _a^{\,\,\,\,\,b} \left[ { - \left( {f_i^I f_{\bar j}^K G^{i\bar j}  + L^I \bar L^K } \right)\theta _{KJ}  + \left( {f_{\bar i}^I f_j^K G^{\bar ij}  + \bar L^I L^K  } \right)\theta _{KJ} } \right. \\
 & & \left. { - i\left( {f_i^I f_{\bar j}^K G^{i\bar j}  + L^I \bar L^K } \right)\gamma _{KJ}  - i\left( {f_{\bar i}^I f_j^K G^{\bar ij}  + \bar L^I L^K  } \right)\gamma _{KJ} } \right]\left( {\partial _b \tilde H^J }
 \right).
\end{eqnarray}

Once again, the terms in $\theta$ cancel out, while, using
(\ref{comp4}), the terms in $\gamma$ add up to:
\begin{equation}
    \left( {\partial _a \zeta ^I } \right) = \frac{i}{{4\sqrt 2 }} \varepsilon _a^{\,\,\,\,\,b} \left( { - i\gamma ^{IK} \gamma _{KJ}} \right)\left( {\partial _b \tilde H^J } \right) = \frac{1}{{4\sqrt 2 }} \varepsilon _a^{\,\,\,\,\,b} \delta _J^I \left( {\partial _b \tilde H^J } \right) = \frac{1}{{4\sqrt 2 }} \varepsilon _a^{\,\,\,\,\,b} \left( {\partial _b \tilde H^I }
    \right),
\end{equation}
which, up to a difference in numerical constants due to different
normalization conventions, is the expression (\ref{zetaeom}) for
$\left( {\partial _a \zeta ^I } \right)$ found using the five
dimensional equations! The calculation for $\left( {\partial _a
\tilde \zeta _I } \right)$ is very similar, and is only different in
the usage of (\ref{comp5}) instead of (\ref{comp4}) in the
appropriate steps, giving:
\begin{equation}
    \left( {\partial _a \tilde \zeta _I } \right) = \frac{1}{{4\sqrt 2 }} \varepsilon _a^{\,\,\,\,\,b} \left( {\partial _b H_I }
    \right).
\end{equation}

Finally, we look at the geometric significance of the constraints
(\ref{const3}, \ref{const4}, \ref{const1}, \ref{const2}). It is
clear that in the far field limit, (\ref{const3}) and
(\ref{const4}) vanish identically. Condition (\ref{const1}), on
the other hand, may be understood as follows: From
(\ref{determinant}), one sees that $g=Hk$, where $k \equiv \det
\left( {k_{\tilde I\tilde J} } \right)$. From (\ref{const1}) one
then concludes that $k=1$, \textit{i.e.} constant, which is simply
the statement that $k_{\tilde I\tilde J}$ is Ricci-flat. For the
last condition (\ref{const2}), we proceed in the following way:
\begin{equation}
    \Omega ^{\tilde I\tilde J\tilde K} \left( {\partial _a \bar \Omega _{\tilde I\tilde J\tilde K} } \right) = 12\left( {\partial _a \ln g} \right) = 12\left( {\partial _a \ln H} \right) =  - 24\left( {\partial _a \sigma }
    \right).
\end{equation}

Since the dilaton is a real field, then this is the statement that
$\Omega ^{\tilde I\tilde J\tilde K} \left( {\partial _a \bar
\Omega _{\tilde I\tilde J\tilde K} } \right)$ is a real quantity.
Then by setting its imaginary part equal to zero, we may proceed
with a bit of algebra as follows:
\begin{eqnarray}
 \Omega ^{\tilde I\tilde J\tilde K} \left( {\partial _a \bar \Omega _{\tilde I\tilde J\tilde K} } \right) = {\mathop{\rm Re}\nolimits} \Omega ^{\tilde I\tilde J\tilde K} \left( {\partial _a {\mathop{\rm Re}\nolimits} \bar \Omega _{\tilde I\tilde J\tilde K} } \right) + {\mathop{\rm Im}\nolimits} \Omega ^{\tilde I\tilde J\tilde K} \left( {\partial _a {\mathop{\rm Im}\nolimits} \bar \Omega _{\tilde I\tilde J\tilde K} } \right) &=&  - 24\left( {\partial _a \sigma } \right) \nonumber\\
 \left[ {\bar \Omega ^{\tilde I\tilde J\tilde K} \left( {\partial _a \Omega _{\tilde I\tilde J\tilde K} } \right) + \Omega ^{\tilde I\tilde J\tilde K} \left( {\partial _a \bar \Omega _{\tilde I\tilde J\tilde K} } \right)} \right] &=&  - 48\left( {\partial _a \sigma }
 \right).
\end{eqnarray}

Since
\begin{equation}
    \left[ {\bar \Omega ^{\tilde I\tilde J\tilde K} \left( {\partial _a \Omega _{\tilde I\tilde J\tilde K} } \right) + \Omega ^{\tilde I\tilde J\tilde K} \left( {\partial _a \bar \Omega _{\tilde I\tilde J\tilde K} } \right)} \right] = \partial _a \left( {\Omega ^{\tilde I\tilde J\tilde K} \bar \Omega _{\tilde I\tilde J\tilde K} } \right) = 3!\left( {\partial _a \left| \Omega  \right|^2 }
    \right),
\end{equation}
then
\begin{equation}
    \left( {\partial _a \sigma } \right) = - \frac{1}{8}\left( {\partial _a \left| \Omega  \right|^2 }
    \right).
\end{equation}

In other words, (\ref{const2}) is simply the statement that the
dilaton field is proportional to the norm of the Calabi-Yau 3-form,
which is a constant on $\mathcal{M}$ but is not necessarily so at
any point in the transverse space, due to the variation of the
complex structure moduli of $\M$.

\section*{Conclusion}

We have explicitly shown that the dimensional reduction of a
M5-brane wrapping special Lagrangian 3-cycles of a Calabi-Yau
manifold deforming the complex structure \cite{Fayyauddin:2005as}
excites only the hypermultiplets sector of ungauged five
dimensional $\mathcal{N}=2$ supergravity. The $D=5$ universal
axion $a$ (or its dual 3-form gauge field) vanishes and the result
is a 2-brane coupled to the hypermultiplet fields
\cite{Emam:2005bh}. This constitutes a proof of this relationship,
often quoted in the literature, as well as further confirmation of
both results. In addition, it provides further confirmation of the
interpretation of \cite{Fayyauddin:2005as} as a wrapped brane.
This paper may be thought of as the sequel to an argument
presented in \cite{Emam:2005bh} where a, much shorter, calculation
has shown that a certain configuration of M-branes wrapping SLAG
cycles of a Calabi-Yau with constant complex structure moduli
excite only the universal hypermultiplet and result in a special
case of the more general five dimensional 2-brane with full
hypermultiplets discussed therein and here.

Along with \cite{Kastor:2003jy}, our calculation provides more hints
to open questions concerning compactification mechanisms, what
classes of Calabi-Yau metrics are relevant and so on. Calculations
such as these may also help analyze other brane configurations
wrapping SLAG-calibrated cycles, as well as other supersymmetric
cycles in spaces with restricted holonomy. For example, it would be
interesting to see what lower dimensional results could arise from
the dimensional reduction of M-branes wrapping manifolds with $G_2$
or $spin(7)$ holonomy. More interesting than the result perhaps, as
often happens, is the manner with which the result arises. We plan to
explore this in future work.

\section*{Acknowledgments}

We are grateful to Ansar Fayyazuddin and David Kastor for pointing
out some subtleties in the calculation and for taking the time to
read the manuscript and providing further advice.


\begin{thebibliography}{999}

\bibitem{Gibbons:1982fy}
  G.~W.~Gibbons and C.~M.~Hull,
   ``A Bogomolny Bound For General Relativity And Solitons In N=2 Supergravity,''
  Phys.\ Lett.\ B {\bf 109}, 190 (1982).

\bibitem{Berger}
  M.~Berger,
   ``Sur les groupes d'holonomie homog\`{e}ne des vari\'{e}t\'{e}s
  \`{a} connexion affine et des vari\'{e}t\'{e}s riemanniennes,''
  Bull.\ Soc.\ Math.\ France {\bf 83}, 225 (1955).

\bibitem{Joyce:2001nm}
  D.~Joyce,
  ``Lectures on Calabi-Yau and special Lagrangian geometry,''
  [arXiv:math.dg/0108088].

\bibitem{Emam:2004nc}
  M.~H.~Emam,
  ``Calibrated brane solutions of M-theory,''
  [arXiv:hep-th/0410100].

\bibitem{Husain:2003ag}
  T.~Z.~Husain,
  ``If I only had a brane!,''
  [arXiv:hep-th/0304143].

\bibitem{Cho:2000hg}
  H.~Cho, M.~Emam, D.~Kastor and J.~H.~Traschen,
  ``Calibrations and Fayyazuddin-Smith spacetimes,''
  Phys.\ Rev.\ D {\bf 63}, 064003 (2001)
  [arXiv:hep-th/0009062].

\bibitem{Fayyazuddin:1999zu}
  A.~Fayyazuddin and D.~J.~Smith,
  ``Localized intersections of M5-branes and four-dimensional  superconformal
  field theories,''
  JHEP {\bf 9904}, 030 (1999)
  [arXiv:hep-th/9902210].

\bibitem{Emam:2005bh}
  M.~H.~Emam,
   ``Five dimensional 2-branes from special Lagrangian wrapped M5-branes,''
  Phys.\ Rev.\ D {\bf 71}, 125020 (2005)
  [arXiv:hep-th/0502112].

\bibitem{Martelli:2003ki}
  D.~Martelli and J.~Sparks,
  ``G-structures, fluxes and calibrations in M-theory,''
  Phys.\ Rev.\ D {\bf 68}, 085014 (2003)
  [arXiv:hep-th/0306225].

\bibitem{Fayyauddin:2005as}
  A.~Fayyazuddin and T.~Z.~Husain,
   ``The geometry of M-branes wrapping special Lagrangian cycles,''
  [arXiv:hep-th/0505182].

\bibitem{Fayyazuddin:2005pm}
  A.~Fayyazuddin, T.~Z.~Husain and I.~Pappa,
  ``The geometry of wrapped M5-branes in Calabi-Yau 2-folds,''
  Phys.\ Rev.\ D {\bf 73}, 126004 (2006)
  [arXiv:hep-th/0509018].

\bibitem{Fayyazuddin:2005ds}
  A.~Fayyazuddin and T.~Z.~Husain,
  ``Calibrations, torsion classes and wrapped M-branes,''
  Phys.\ Rev.\ D {\bf 73}, 106007 (2006)
  [arXiv:hep-th/0512030].

\bibitem{Kastor:2003jy}
  D.~Kastor,
  ``From wrapped M-branes to Calabi-Yau black holes and strings,''
  JHEP {\bf 0307}, 040 (2003)
  [arXiv:hep-th/0305261].

\bibitem{Cadavid:1995bk}
A.~C.~Cadavid, A.~Ceresole, R.~D'Auria and S.~Ferrara,
``Eleven-dimensional supergravity compactified on Calabi-Yau
threefolds,'' Phys.\ Lett.\ B {\bf 357}, 76 (1995)
[arXiv:hep-th/9506144].

\bibitem{Candelas:1990pi}
P.~Candelas and X.~de la Ossa, ``Moduli Space Of Calabi-Yau
Manifolds,'' Nucl.\ Phys.\ B {\bf 355}, 455 (1991).

\bibitem{Gutperle:2000ve}
M.~Gutperle and M.~Spalinski, ``Supergravity instantons for N = 2
hypermultiplets,'' Nucl.\ Phys.\ B {\bf 598}, 509 (2001)
[arXiv:hep-th/0010192].

\bibitem{DeJaegher:1997ka}
J.~De Jaegher, B.~de Wit, B.~Kleijn and S.~Vandoren, ``Special
geometry in hypermultiplets,'' Nucl.\ Phys.\ B {\bf 514}, 553
(1998) [arXiv:hep-th/9707262].

\bibitem{Witten:mh}
E.~Witten and D.~I.~Olive, ``Supersymmetry Algebras That Include
Topological Charges,'' Phys.\ Lett.\ B {\bf 78}, 97 (1978).

\bibitem{Sabra:1997dh}
  W.~A.~Sabra,
  ``Black holes in N = 2 supergravity theories and harmonic functions,''
  Nucl.\ Phys.\ B {\bf 510}, 247 (1998)
  [arXiv:hep-th/9704147].

\end{thebibliography}
\end{document}